\documentstyle[preprint,epsf,aps,floats,tighten]{revtex} 

\input{psfig}

\def\met{\mbox{${\hbox{$E$\kern-0.6em\lower-.1ex\hbox{/}}}_T~$}} 
\def\D0{D\O}                            
\begin{document}

\preprint{FERMILAB-Conf-98/279-E}
%
%
\title{Coloron Limits using the D\O\ Dijet Angular Distribution}

\author{\centerline{The D\O\ Collaboration
  \thanks{Submitted to the {\it XXIX International Conference on
          High Energy Physics - ICHEP98,}
          \hfill\break
          July 23 -- 29, 1998, Vancouver, B.C., Canada.}}}
\address{
\centerline{Fermi National Accelerator Laboratory, Batavia, Illinois 60510}
}

%
%
\date{\today}

\maketitle

%
%
\begin{abstract}
The D\O\ dijet angular distribution is used to place limits on
flavor--universal colorons. Models with $M_c / \cot{\theta} < 759$
GeV/c$^2$ are excluded at the 95$\%$ confidence level.
\end{abstract}

\newpage
\begin{center}
%
B.~Abbott,$^{40}$                                                             
M.~Abolins,$^{37}$                                                            
V.~Abramov,$^{15}$                                                            
B.S.~Acharya,$^{8}$                                                           
I.~Adam,$^{39}$                                                               
D.L.~Adams,$^{48}$                                                            
M.~Adams,$^{24}$                                                              
S.~Ahn,$^{23}$                                                                
H.~Aihara,$^{17}$                                                             
G.A.~Alves,$^{2}$
% ...
\\                                                                            
\vskip 0.50cm                                                                 
\centerline{(D\O\ Collaboration)}                                             
\vskip 1.0cm
\centerline{$^{1}$Universidad de Buenos Aires, Buenos Aires, Argentina}       
\centerline{$^{2}$LAFEX, Centro Brasileiro de Pesquisas F{\'\i}sicas,         
                  Rio de Janeiro, Brazil}                                     
\centerline{$^{3}$Universidade do Estado do Rio de Janeiro,                   
                  Rio de Janeiro, Brazil}                                     
\centerline{$^{4}$Universidad de los Andes, Bogot\'{a}, Colombia}             
\centerline{$^{5}$DAPNIA/Service de Physique des Particules, CEA, Saclay,     
                  France}                                                     
\centerline{$^{6}$Panjab University, Chandigarh, India}                       
\centerline{$^{7}$Delhi University, Delhi, India}                             
\centerline{$^{8}$Tata Institute of Fundamental Research, Mumbai, India}      
\centerline{$^{9}$Kyungsung University, Pusan, Korea}                         
\centerline{$^{10}$Seoul National University, Seoul, Korea}                   
\centerline{$^{11}$CINVESTAV, Mexico City, Mexico}                            
\centerline{$^{12}$Institute of Nuclear Physics, Krak\'ow, Poland}            
\centerline{$^{13}$Institute for Theoretical and Experimental Physics,        
                   Moscow, Russia}                                            
\centerline{$^{14}$Moscow State University, Moscow, Russia}                   
\centerline{$^{15}$Institute for High Energy Physics, Protvino, Russia}       
\centerline{$^{16}$University of Arizona, Tucson, Arizona 85721}              
\centerline{$^{17}$Lawrence Berkeley National Laboratory and University of    
                   California, Berkeley, California 94720}                    
\centerline{$^{18}$University of California, Davis, California 95616}         
\centerline{$^{19}$University of California, Irvine, California 92697}        
\centerline{$^{20}$University of California, Riverside, California 92521}     
\centerline{$^{21}$Florida State University, Tallahassee, Florida 32306}      
\centerline{$^{22}$University of Hawaii, Honolulu, Hawaii 96822}              
\centerline{$^{23}$Fermi National Accelerator Laboratory, Batavia,            
                   Illinois 60510}                                            
\centerline{$^{24}$University of Illinois at Chicago, Chicago,                
                   Illinois 60607}                                            
\centerline{$^{25}$Northern Illinois University, DeKalb, Illinois 60115}      
\centerline{$^{26}$Northwestern University, Evanston, Illinois 60208}         
\centerline{$^{27}$Indiana University, Bloomington, Indiana 47405}            
\centerline{$^{28}$University of Notre Dame, Notre Dame, Indiana 46556}       
\centerline{$^{29}$Purdue University, West Lafayette, Indiana 47907}          
\centerline{$^{30}$Iowa State University, Ames, Iowa 50011}                   
\centerline{$^{31}$University of Kansas, Lawrence, Kansas 66045}              
\centerline{$^{32}$Louisiana Tech University, Ruston, Louisiana 71272}        
\centerline{$^{33}$University of Maryland, College Park, Maryland 20742}      
\centerline{$^{34}$Boston University, Boston, Massachusetts 02215}            
\centerline{$^{35}$Northeastern University, Boston, Massachusetts 02115}      
\centerline{$^{36}$University of Michigan, Ann Arbor, Michigan 48109}         
\centerline{$^{37}$Michigan State University, East Lansing, Michigan 48824}   
\centerline{$^{38}$University of Nebraska, Lincoln, Nebraska 68588}           
\centerline{$^{39}$Columbia University, New York, New York 10027}             
\centerline{$^{40}$New York University, New York, New York 10003}             
\centerline{$^{41}$University of Rochester, Rochester, New York 14627}        
\centerline{$^{42}$State University of New York, Stony Brook,                 
                   New York 11794}                                            
\centerline{$^{43}$Brookhaven National Laboratory, Upton, New York 11973}     
\centerline{$^{44}$University of Oklahoma, Norman, Oklahoma 73019}            
\centerline{$^{45}$Brown University, Providence, Rhode Island 02912}          
\centerline{$^{46}$University of Texas, Arlington, Texas 76019}               
\centerline{$^{47}$Texas A\&M University, College Station, Texas 77843}       
\centerline{$^{48}$Rice University, Houston, Texas 77005}                     

\end{center}

\normalsize

\vfill\eject
 A flavor universal coloron model~\cite{coloron_1} has been proposed
 to explain the excess in the inclusive jet cross section as measured
 by CDF~\cite{cdf_jet} without contradicting other experimental
 data. The model is minimal in its structure, in that it involves the
 addition of one new interaction, one new scalar multiplet and no new
 fermions. The QCD gauge group is extended to $SU(3)_{1} \times
 SU(3)_{2}$, with gauge couplings $\xi_{1}$ and $\xi_{2}$
 respectively, with $\xi_2 \gg \xi_1$. All quarks are assigned to
 triplet representations of the stronger group $SU(3)_{2}$. The
 symmetry is broken to its diagonal subgroup, $SU(3)_{\rm QCD}$, when
 a scalar boson $\Phi$ which transforms as a (3,$\bar{3}$) under the
 original two SU(3) groups acquires a vacuum expectation value
 $<\Phi>= f$.  Thus at low energies there exist both ordinary massless
 gluons and an octet of heavy coloron bosons, $C^{\mu a}$.  As described
 in ~\cite{coloron_1}, the colorons couple to all quarks as:
\begin{equation}
{\mathcal{L}} = -g_{3} \cot{\theta} J_{\mu}^{a} C^{\mu a},
\end{equation}
 where $J_{\mu}^{a}$ is the color current ($J_{\mu}^{a}=\bar{q}
 \gamma^\mu (\lambda^{a}/2) q$) and $\cot{\theta} = \xi_{2} /
 \xi_{1}$. The coupling $g_3$ is identified with the QCD coupling
 constant and has a value of approximately 1.2 (corresponding to
 $\alpha_{3}(M_{Z}) \approx 0.12$). In terms of these parameters the
 mass of the coloron is
\begin{equation}
M_{c} = \left( {{g_3} \over {\sin{\theta} \cos{\theta}}} \right) f.
\end{equation}
 Below the scale $M_c$, coloron--exchange can be approximated by the
 effective four--fermion interaction
\begin{equation}
{\mathcal{L}}_{\rm eff} = 
- { {g_{3}^{2} \cot^{2}{\theta}} \over {2! M_{c}^{2}}} 
J_{\mu}^{a} J^{\mu a}.
\label{eqn_3}
\end{equation}
 Note, this interaction can be rewritten in the form usually used in
 studies of quark compositeness:
\begin{equation}
{\mathcal{L}}_{\rm eff} = {{- g_3^2}\over{2! \Lambda^2}} J^a_\mu J^{a\mu}  
\end{equation}
 where $g_3^2 \equiv 4\pi$ and $\Lambda \sqrt{\alpha_{\rm s}} =
 M_{c}/\cot{\theta}$.

 The phenomenology of the coloron has been studied~\cite{coloron_2}
 and limits have been placed on $M_c$ and $\cot{\theta}$. Constraints
 on the size of the weak--interaction $\rho$ parameter require
 \mbox{$M_{c} / \cot{\theta} > 450$~GeV}~\cite{coloron_1} and a direct
 search for colorons in the dijet mass spectrum at CDF~\cite{cdfxjj}
 excludes colorons with mass below 1 TeV for $\cot{\theta} \lesssim
 1.5$.

\begin{figure}[hbtp]
\epsfxsize=4.5in
\centerline{\epsffile{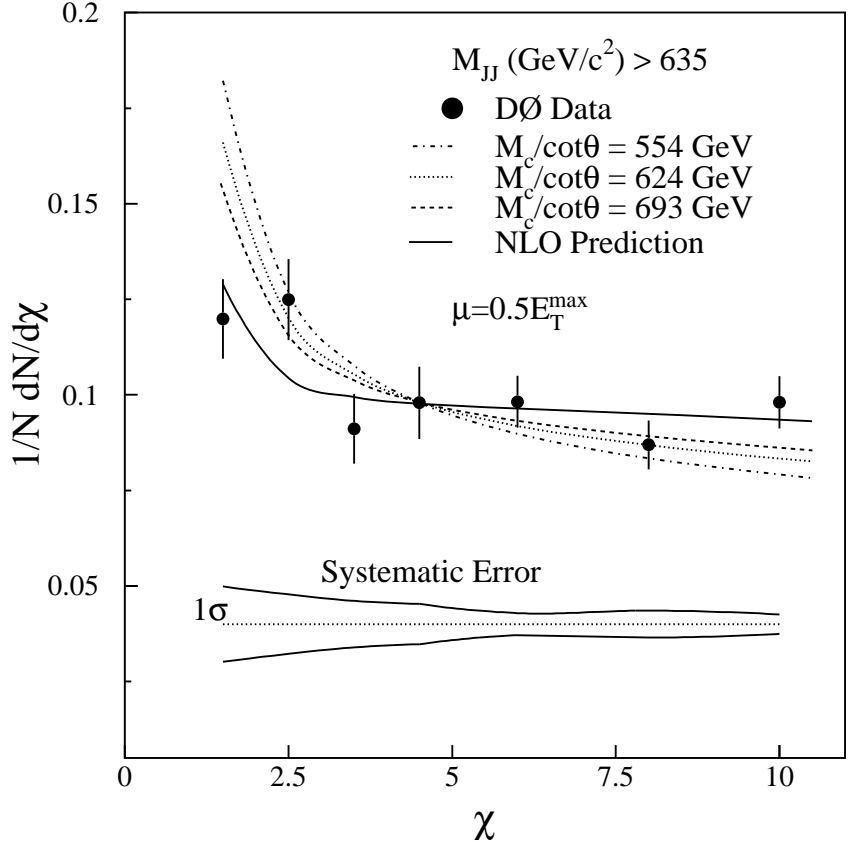}}
\caption{D\O\ data compared to theory for different values of 
 \mbox{$M_{c} / \cot{\theta}$} (see text for details on how the
 coloron distributions are calculated). The errors on the points are
 statistical and the band represents the correlated $\pm 1\sigma$
 systematic uncertainty.}
\label{fig_1}
\end{figure}

 In analogy with the effects of quark compositeness~\cite{lane_1}, the
 low-energy effects of coloron exchange (equation~\ref{eqn_3}) will
 lead to to an excess of events at small values of $\chi$ in the dijet
 angular distributions ($\chi = \exp ( \Delta \eta )$ where $\Delta
 \eta$ is the separation in pseudorapidity between the two jets ($\eta
 = -{\rm ln}[{\rm tan}(\theta/2)]$)). Predictions of the dijet angular
 distribution with colorons are available at leading order (LO). To
 simulate a next--to--leading order (NLO) prediction, the effects of
 the coloron LO predictions for several values of \mbox{$M_{c} /
 \cot{\theta}$} are generated. The fractional difference between the
 angular distribution with \mbox{$M_{c} / \cot{\theta} = \infty$} and
 the distribution with a finite value of \mbox{$M_{c} / \cot{\theta}$}
 are then measured. The coloron NLO prediction is then obtained by
 multiplying a NLO QCD prediction obtained using the {\sc Jetrad}
 program~\cite{jetrad} by the LO fractional differences obtained
 above. The results are shown in Fig.~\ref{fig_1}.

 Also shown in Fig.~\ref{fig_1} is the dijet angular distribution as
 measured by D\O~\cite{d0_ang} for \mbox{$M_{\rm JJ} > 635$
 GeV/c$^2$}. Limits on the coloron mass are calculated using the ratio 
\begin{equation}
R_{\chi} = {{N(\chi < 4.) } \over { N ( 4. < \chi < \chi_{\rm max})}},
\end{equation}
 where $N( \chi )$ gives the number of events in the given $\chi$
 range. The calculation of $R_{\chi}$ removes the correlated errors
 that are a function of $\chi$. Table~\ref{table_1} gives $R_{\chi}$
 for the different mass ranges and their statistical and systematic
 uncertainties, which are fully correlated in mass.

\begin{table}[htb]
\begin{center}
\begin{minipage}[t]{4in}
\begin{tabular}{cccc}
\multicolumn{1}{c}{Mass Range (GeV/c$^2$)} &
\multicolumn{1}{c}{$R_{\chi}$} &
\multicolumn{1}{c}{Stat Error} &
\multicolumn{1}{c}{Sys Error} \\ \hline
260-425 & 0.191 & 0.0077 & 0.015 \\
425-475 & 0.202 & 0.0136 & 0.010 \\
475-635 & 0.342 & 0.0085 & 0.018 \\
635-    & 0.506 & 0.0324 & 0.028 \\
\end{tabular}
\end{minipage}
\caption{The dijet angular ratio $R_{\chi}$ and its statistical and
systematic uncertainty.}
\label{table_1}
\end{center}
\end{table}

 The coloron limit is calculated using Bayesian techniques with a
 Gaussian likelihood function ($P(x)$) for $R_{\chi}$ as a function of
 mass:
\begin{equation}
 P(x) = {{1} \over { \mid \! S \! \mid 2 \pi^{2}}} 
\exp{\left( -\frac{1}{2} \left[ d - f(x) \right]^{T} S^{-1} 
\left[ d - f(x) \right] \right)}
\end{equation}
 where $d$ is the vector of data points for the different mass bins,
 $f(x)$ is vector of theory points for the different masses at
 different values of $x$ where $x = 1/\Lambda^{n}$, and $S^{-1}$ is the
 inverse of the covariance matrix.

\begin{figure}[hbtp]
\epsfxsize=4.5in
\centerline{\epsffile{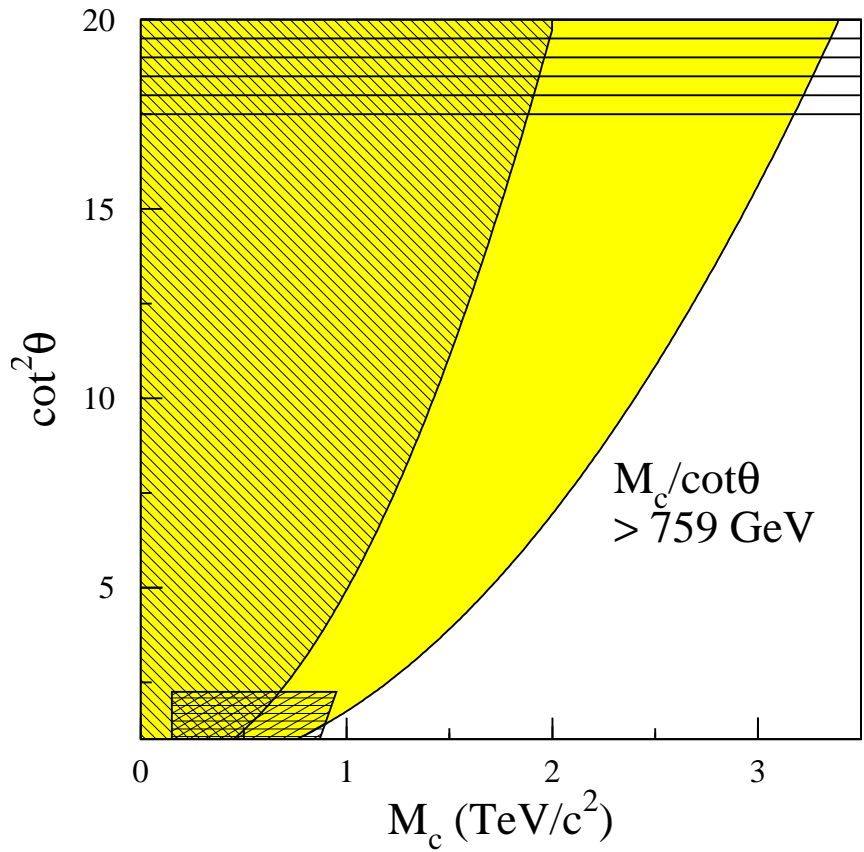}}
\caption{
Limits on the coloron parameter space: coloron mass $M_c$ vs.
mixing parameter $\cot{\theta}$. The shaded region shows the 95$\%$
CL exclusion region for the D\O\ dijet angular distribution
measurement (\mbox{$M_{c} / \cot{\theta} > 759$ GeV/c$^2$}). The
horizontally hatched region at large $\cot{\theta}$ is not allowed in
this phase of the model~[1,3]. The
diagonally hatched region is excluded by the value of $\rho$
(\mbox{$M_{c} / \cot{\theta} > 450$ GeV/c$^2$})~[1]. The
cross--hatched region is excluded by the CDF search for new particles
decaying to dijets~[4].
}
\label{fig_2}
\end{figure}

 The limit depends on the choice of prior probability distribution for
 $P(x)$. Motivated by the form of the Lagrangian the prior probability
 is assumed to be flat when $x = 1/\Lambda^{2}$. Since the dijet
 angular distribution is sensitive to the choice of renormalization
 scale, each possible choice is treated as a different theory. The
 95$\%$ confidence limit (CL) on $\Lambda$ is calculated by requiring that
\begin{equation}
Q(x) = \int_{0}^{x}  P(x) dx = 0.95 Q(\infty).
\end{equation}
 The limit in $x$ is then transformed back into a limit on
 $\Lambda$. If a renormalization scale, $\mu = E_{T}^{\rm max}$ (where
 $E_{T}^{\rm max}$ is the transverse energy of the highest $E_T$ jet)
 then the 95$\%$ CL on the coloron mass is \mbox{$M_{c} / \cot{\theta}
 > 759$ GeV/c$^2$}. If $\mu = 0.5E_{T}^{\rm max}$ then the 95$\%$ CL
 is \mbox{$M_{c} / \cot{\theta} > 786$ GeV/c$^2$}. The resulting
 limits are plotted in Fig~\ref{fig_2}.

 In conclusion, the dijet mass spectrum as measured by D\O\ can be
 used to exclude flavor--universal colorons with \mbox{$M_{c} /
 \cot{\theta}$} below 759 GeV/c$^{2}$ at the 95$\%$ confidence level.

\section*{Acknowledgements}
\label{sec:ack}
%
%

We thank R. Harris for the use of his program based on
Refs. \cite{lane_1,coloron_1}. We thank the staffs at Fermilab and
collaborating institutions for their contributions to this work, and
acknowledge support from the Department of Energy and National Science
Foundation (U.S.A.), Commissariat \` a L'Energie Atomique (France),
State Committee for Science and Technology and Ministry for Atomic
Energy (Russia), CAPES and CNPq (Brazil), Departments of Atomic Energy
and Science and Education (India), Colciencias (Colombia), CONACyT
(Mexico), Ministry of Education and KOSEF (Korea), and CONICET and
UBACyT (Argentina).

\end{document}